\documentclass[%
 reprint,
 amsmath,amssymb,
 aps,
]{revtex4-2}
\usepackage{color}
\usepackage{graphicx}
\usepackage{dcolumn}
\usepackage{bm}

\begin{document}

\preprint{APS/123-QED}

\title{Eulerian-Lagrangian scaling of the Lyapunov exponent in
	homogeneous turbulence}

\author{Jin GE}
 \email{jin.ge@cnrs.fr}
\author{Joran ROLLAND}%
 \email{joran.rolland@centralelille.fr}
 \author{John Christos VASSILICOS}%
 \email{john-christos.vassilicos@cnrs.fr}
\affiliation{%
 Univ. Lille, CNRS, ONERA, Arts et Métiers ParisTech,
 Centrale Lille, UMR 9014 - LMFL - Laboratoire de Mécanique des
 Fluides de Lille - Kampé de Feriet, F-59000 Lille, France
}%

\date{\today}

\begin{abstract}
We present a heuristic derivation of the maximal Lyapunov exponent
$\gamma$ of homogeneous turbulence which yields two new relations, a
sweeping relation and the scaling of the uncertainty field's integral
length $L_{\Delta}$. These relations and the maximal Lyapunov
exponent's scaling
that they imply are confirmed by periodic turbulence simulations.
As the Reynolds number $Re_{\lambda}$ increases,
$L_{\Delta}$ and $\gamma^{-1}$ decrease towards values smaller than
the Kolmogorov length and time scales.
\end{abstract}

\maketitle


\section{\label{sec:Introduction}Introduction}

In the presence of chaos and strange attractors, non-linear
deterministic systems are extremely sensitive to the smallest
uncertainties in initial conditions. This sensitivity, commonly known
as the ``butterfly effect'', severely limits their predictability even
when their evolution equations are perfectly well known. Lorenz
\cite{lorenz1963deterministic} demonstrated this effect on a system of
three non-linear differential equations derived from an
oversimplification of thermal convection equations. However, the
butterfly effect is also present in more realistic situations with a
much higher number of degrees of freedom such as in
magnetohydrodynamics \cite{ho2020fluctuations}, turbulent convection
and turbulent flows in general
\cite{deissler1986navier,mohan2017scaling,boffetta2017chaos,berera2018chaotic},
weather forecasting and atmospheric predictability
\cite{leith1971atmospheric,palmer2000predicting}. The presence of the
butterfly effect in fluid turbulence governed by the Navier-Stokes
equations was first demonstrated by Deissler \cite{deissler1986navier}
in 1986.

In all these cases, estimating the maximal Lyapunov exponent in terms
of the parameters of the specific system
is an important and challenging question. 
In this letter the focus is on turbulent flow which is a generic case
of deterministic chaotic non-linear system with a very large number of
degrees of freedom. For such a system, Ruelle
\cite{ruelle1979microscopic} argued that the time-scale governing
chaos (the inverse maximal Lyapunov exponent) is proportional to the
smallest Lagrangian time scale of the turbulence irrespective of
Reynolds number $Re_{\lambda}$.
Could the butterfly effect act over a time scale that tends to become
even smaller than this smallest Lagrangian time scale with increasing
$Re_{\lambda}$, and if so why?  We answer this question by arguing
that it is not enough to consider the smallest Lagrangian time scale
as done by Ruelle, but that the smallest Eulerian time scale should
also be taken into account. This requires consideration of the spatial
structure of the uncertainty field and of its random sweeping by
turbulent eddies.

\section{\label{sec:Theoretical approach}Theoretical approach}

A tiny small-scale uncertainty in a
turbulent velocity field $\bm{u}$ gives rise to a turbulent velocity
field $\bm{u} + \Delta\bm{u}$ which quickly decorrelates from
$\bm{u}$. Specifically, for statistically stationary and homogeneous
turbulence, the space-average energy $\left\langle
E_{\Delta}\right\rangle\equiv\left\langle\left|\Delta\bm{u}\right|^{2}/2\right\rangle$
of the uncertainty field $\Delta\bm{u} (\bm{x},t)$ evolves according
to
\begin{equation}
	\label{eq:exponential growth of uncertainty energy}
	\frac{\mathrm{d}}{\mathrm{d}t}\left\langle E_{\Delta}\right\rangle=2\gamma\left\langle E_{\Delta}\right\rangle,
\end{equation}  
where $\left\langle\cdot\right\rangle$ represents average over spatial
positions $\bm{x}$, and where $\gamma$ is the maximal Lyapunov
exponent. This chaotic behaviour is a relatively early time behaviour
which lasts till times $t$ when the fields $\bm{u}$ and $\bm{u} +
\Delta\bm{u}$ have fully decorrelated only at the smallest resolvable
length scales \cite{ge2023production}. A fundamental property of the
growth of uncertainty in turbulence is the scaling of $\gamma$ with
Reynolds number.
A Reynolds number $Re_{\lambda}$ can be defined as the ratio of the
turbulent kinetic energy $3U^{2}/2$
to the kinetic energy $\sqrt{\nu \varepsilon}$ at the Kolmogorov
length $\eta\equiv\left(\nu^{3}/\varepsilon\right)^{1/4}$:
$Re_{\lambda} = 3U^{2}/(2\sqrt{\nu \varepsilon})$ ($\nu$ is the
fluid's kinematic viscosity and $\varepsilon$ is the turbulence
dissipation rate).


Ruelle argued that when two statistically stationary homogeneous
turbulent velocity fields, $\bm{u}$ and $\bm{u} + \Delta\bm{u}$,
differ initially only at the smallest scales, $\gamma^{-1}$ should
correspond to the shortest time scale in the turbulence, and picked
the Kolmogorov time scale
$\tau_{\eta}\equiv\left(\nu/\varepsilon\right)^{1/2}$
\cite{ruelle1979microscopic}. Hence, Ruelle's prediction is
$\gamma\tau_{\eta} \sim 1$.
However, Ruelle's argument does not distinguish between Eulerian and
Lagrangian time scales. Whereas $\tau_{\eta}$ is the smallest
Lagrangian time scale, $\tau_{E}\equiv\eta/U$ is the smallest Eulerian
time scale resulting from random sweeping of the smallest eddies by
the largest ones \cite{tennekes1975eulerian}.

Here we propose a heuristic derivation of the $Re_{\lambda}$ scaling
of $\gamma \tau_{\eta}$ for statistically stationary homogeneous
isotropic turbulence which takes into account both the Lagrangian and
the Eulerian smallest time scales. As demonstrated in
  \cite{ge2023production}, during the chaotic uncertainty growth when
  Eq. (\ref{eq:exponential growth of uncertainty energy}) holds, the
  uncertainty is highly amplified only in a relatively small number of
  locations in the flow. Our derivation is founded on the hypothesis
  that these relatively sparse locations are in the vicinity of
  stagnation points of the fluctuating reference field ${\bf u}({\bf
    x}, t)$. The most significant part of the fluctuations of the
  uncertainty field $\Delta {\bf u}$, which are small-scale
  low-amplitude fluctuations, are therefore concentrated around these
  stagnation points where the uncertainty or noise to reference signal
  ratio is the highest. Naturally, one can expect these small-scale
  uncertainty fluctuations to be swept by reference field eddies
  larger than them.
Random sweeping therefore influences uncertainty growth
which, as shown below, leads to $\gamma\tau_{\eta} \sim
Re_{\lambda}^{1/3}$, meaning that $\gamma$ grows faster than
$\tau_{\eta}^{-1}$ but slower than $\tau_{E}^{-1} \sim
\tau_{\eta}^{-1} Re_{\lambda}^{1/2}$ with increasing $Re_{\lambda}$.



The sweeping part of our model is formulated in terms of the
uncertainty-containing length
$L_{\Delta}\equiv\left(3\pi/4\left\langle
E_{\Delta}\right\rangle\right)\int k^{-1}\hat{E}_{\Delta}(k){\rm d}k$,
where $\hat{E}_{\Delta}(k)$ is the uncertainty field's energy spectrum
\cite{ge2023production}. $L_{\Delta}$ is the integral length scale
\cite{batchelor1953theory} of the uncertainty field. It has been shown
\cite{ge2023production} to be constant in time for statistically
stationary homogeneous isotropic turbulence during chaotic exponential
growth of uncertainty.
Random sweeping trivially vanishes at reference field stagnation
points given the zero fluctuating velocity at these points. The
fluctuating uncertainty field around these points is swept by
reference field velocities which are therefore typically small and
dominated by eddies of average size equal to the average distance
$\overline{l_{e}}$ between reference field stagnation points. We
therefore expect uncertainty fluctuations to be randomly swept by
reference field eddies of size $\overline{l_{e}}$ and RMS fluctuating
velocity $U_{\bar{l_e}}\equiv\left(\varepsilon\bar{l_e}\right)^{1/3}$.
It will therefore take a time $\sim
  \frac{L_{\Delta}}{U_{\bar{l_e}}}$ for the uncertainty fluctuations
  to be swept away from the region around stagnation points where they
  are being amplified (taking into account that positions of
  stagnation points are quite persistent as demonstrated by
  \cite{Persist2005}). The time duration of this amplification may
  then be $\frac{L_{\Delta}}{U_{\bar{l_e}}}$ and it may be natural to
  expect it to be of the same order as the Lyapunov time
  $\tau_{\gamma}\equiv \gamma^{-1}$ which is the inverse rate of
  amplification. Hence,
\begin{equation}
	\label{eq:assumption advect 2 product}
	\frac{L_{\Delta}}{U_{\bar{l_e}}}\sim\gamma^{-1}. 
\end{equation}

It is known theoretically and computationally for homomogeneous
isotropic turbulence \cite{goto2009dissipation} that the average
distance between stagnation points scales as $\lambda$, the Taylor
length.
Hence $\overline{l_e} \sim \lambda$, and (\ref{eq:assumption advect 2 product}) becomes
\begin{equation}
	\label{eq:assumption advect 2 product 2}
	\gamma L_{\Delta}/U\sim Re_{\lambda}^{-1/3}
\end{equation}
if use is made of $\varepsilon = 15\nu U^{2}/\lambda^{2}$
\cite{taylor1935statistical}.

To derive the $Re_{\lambda}$ scaling of $\gamma$ from
(\ref{eq:assumption advect 2 product 2}) we need the $Re_{\lambda}$
scaling of $L_{\Delta}$. We start with velocity component fluctuations
along one-dimensional straight line cuts through the
turbulence. Because of isotropy, the average distance between
zero-crossings of this velocity component scales in the same way as
the average distance between stagnation points in the
three-dimensional flow (a theoretical result confirmed by experimental
and computational data, see
\cite{goto2009dissipation,mazellier2008turbulence}). Denoting by $l$
the random distance between consecutive zero-crossings and by
$\overline{l}$ the average distance between them, the Rice theorem
\cite{mazellier2008turbulence,rice1945mathematical,liepmann1953counting}
implies $\overline{l} \sim \lambda$ for all Reynolds
numbers. (Dissipation range intermittency introduces
  a small $\log Re_{\lambda}$ correction to this scaling, see
  \cite{mazellier2008turbulence}, which we do not take into account
  here.)
  
We then formalise our hypothesis that, during the chaotic uncertainty
growth, the uncertainty field fluctuations are concentrated around
zero-crossings of one velocity component of the velocity field $\bm{u}
+ \Delta\bm{u}$ on the one-dimensional straight line. It is natural to
assume that there is no characteristic length between such consecutive
zero-crossings. We therefore take (in agreement with
  existing observations
  \cite{davilaprl2003,mazellier2008turbulence,ferran2023characterising})
the probability density function (PDF) of distances $l$ between such
consecutive zero-crossings to be a power law $P(l)=Al^{-q}$ in a range
$l_{\min} \le l \le l_{\max}$ which widens with increasing
$Re_{\lambda}$. (The range $l_{\min}$ to $l_{\max}$ may be thought of
as an inertial range but in what follows we only need
$l_{\max}/l_{\min}$ to grow indefinitely with increasing
$Re_{\lambda}$.) We write
\begin{equation}
	\label{eq:probability=1}
	\mathbb{P}_{l<l_{\min}}+\mathbb{P}_{l>l_{\max}}+\int_{l_{\min}}^{l_{\max}}P(l)\mathrm{d}l=1,
\end{equation}
where $\mathbb{P}_{l<l_{\min}}$ and $\mathbb{P}_{l>l_{\max}}$
represent the probabilities of finding consecutive zero-crossings
separated by a distance below $l_{\min}$ for $\mathbb{P}_{l<l_{\min}}$
and above $l_{\max}$ for
$\mathbb{P}_{l>l_{\max}}$. Given that the early-time
  uncertainty field consists of only very small length scales
  comparable to the smallest length scales in the reference field, we
  expect $P(l)$ to be dominated by zero-crossings of the reference
  field in the range $l_{min}<l<l_{max}$ and
  $\mathbb{P}_{\Delta}\equiv \mathbb{P}_{l<l_{\min}}$ to be dominated
  by zero-crossings of the uncertainty velocity
  fluctuations. Furthermore, we expect the zero-crossings attributable
  to the uncertainty field to be localised around zero-crossings of
  the reference field. Hence, there should be an order one number of
  such localised concentrations of uncertainty zero-crossings within
  an average distance $\overline{l}$ between reference field zero
  crossings.
In a line segment of size $\mathcal{L}$ (e.g. size of the periodic box
in a direct numerical simulation (DNS) of periodic turbulence), there
is, therefore, a number $N_{\Delta} \sim
\mathcal{L}/\bar{l}\sim\mathcal{L}/\lambda$ of localised small
uncertainty fluctuation regions.

The overall length of the entirety of these small
  localised uncertainty fluctuation regions within the line segment of
  size $\mathcal{L}$ can be estimated in two ways: roughly scaling as
  $N_{\Delta}L_{\Delta}$ given that the size of each one of them would
  scale with $L_{\Delta}$, and also as $\mathcal{L}
  \mathbb{P}_{\Delta}$. We can therefore write $N_{\Delta}L_{\Delta}
  \sim \mathcal{L} \mathbb{P}_{\Delta}$ which, combined with
  $N_{\Delta} \sim\mathcal{L}/\lambda$, yields
\begin{equation}
	\label{eq:probability Delta}
	L_{\Delta}\sim\lambda\mathbb{P}_{\Delta}.
\end{equation}
The task of finding the $Re_{\lambda}$ scaling of $L_{\Delta}$ is now
a task of finding the $Re_{\lambda}$ scaling of
$\mathbb{P}_{\Delta}$. We expect $\mathbb{P}_{\Delta}\to 0$ as
$Re_{\lambda}\to \infty$.

From $P(l)=Al^{-q}$, Eq. (\ref{eq:probability=1}) and setting
$\mathbb{P}_{l>l_{\max}}=0$, we obtain
\begin{equation}
	\label{eq:probability lmin-lmax}
	P(l)=\alpha^{-1}(q-1)(1-\mathbb{P}_{\Delta})
	l_{\min}^{-1}\left(l/l_{\min}\right)^{-q},
\end{equation}
where $\alpha=1-\left(l_{\max}/l_{\min}\right)^{1-q}$.
We checked that a choice of fast decreasing $l$-dependence for
$l>l_{\max}$ different from $\mathbb{P}_{l>l_{\max}}=0$ (e.g. an
exponential decrease) does not change our conclusions.


To evaluate $\mathbb{P}_{\Delta}$, we consider equation
(\ref{eq:probability lmin-lmax}) at $l=l_{\min}$, which yields
\begin{equation}
	\label{eq:probability 1-lmin+lmax}
	1-\mathbb{P}_{\Delta}
	=\left[1-\left(\frac{l_{\max}}{l_{\min}}\right)^{1-q}\right]\frac{P(l_{\min})l_{\min}}{q-1}.
\end{equation}
One expects $P(l_{\min})l_{\min}$ to be fully determined by the
smallest turbulence length scales and therefore independent of
$Re_{\lambda}$ at high enough $Re_{\lambda}$, and so, as
$Re_{\lambda}\to+\infty$, Eq. (\ref{eq:probability 1-lmin+lmax}),
$l_{\max}/l_{\min}\to \infty$ and $\mathbb{P}_{\Delta}\to 0$ imply
$q>1$ and
\begin{equation}
	\label{eq:probability lmin+lmax}
	\mathbb{P}_{\Delta}=\left(l_{\max}/l_{\min}\right)^{1-q}.
\end{equation}
We now make use of the Rice theorem $\bar{l}\sim\lambda$ (see
\cite{mazellier2008turbulence}) and of a scaling of the variance of
consecutive zero-crossing lengths $l$ obtained
\cite{mazellier2008turbulence} from data analysis of various types of
turbulence (including turbulent jets, various types of grid-generated
turbulence, and ``chunk'' turbulence in the Modane S1 wind tunnel):
$l_{v}^{2}\equiv \overline{\left(l-\bar{l}\right)^{2}}\sim\lambda^{2}
Re_{\lambda}^{2/3}$.
Writing $l_{v}^{2}=\overline{l^{2}}-\overline{l}^{2}$ and calculating
the integrals $\bar{l} = \int_{0}^{\infty} l P(l)dl$ and $\bar{l^{2}}
= \int_{0}^{\infty} l^{2} P(l)dl$ by neglecting the small contribution
to these integrals coming from $l<l_{\min}$ (one can formally show that
they give rise to neglible higher order corrections) leads to
\begin{equation}
	\label{eq:l2 bar 2 l bar first}
	\frac{l_{v}^{2}}{\overline{l}^{2}}\sim
	\left(\frac{l_{\max}}{l_{\min}}\right)^{q-1}
\end{equation}
for $Re_{\lambda}\gg 1$. The Rice theorem $\overline{l}\sim \lambda$,
equations (\ref{eq:probability lmin+lmax}) and (\ref{eq:l2 bar 2 l bar
	first}), and $l_{v}^{2}\sim\lambda^{2} Re_{\lambda}^{2/3}$ yield
$\mathbb{P}_{\Delta}\sim Re_{\lambda}^{-2/3}$. Hence, from
Eq. (\ref{eq:probability Delta})
\begin{equation}
	\label{eq:LDelta}
	L_{\Delta}\sim\lambda Re_{\lambda}^{-2/3}
\end{equation}
for large enough $Re_{\lambda}$. This result does not require any
specific information on $q$, $l_{\min}$ and $l_{\max}$ except that
$l_{\max}/l_{\min}$ should tend to infinity as $Re_{\lambda}\to
\infty$.

Combining Eq. (\ref{eq:LDelta}) with our sweeping relation
Eq. (\ref{eq:assumption advect 2 product 2}), we finally get the high
Reynolds number scaling of the Lyapunov exponent:
\begin{equation}
	\label{eq:Lyapunov exponent scaling}
	\gamma\lambda/U=\gamma\tau_{\eta}\sim Re_{\lambda}^{1/3}.
\end{equation}
This is equivalent to $\gamma \tau_{\eta}^{1/3}\tau_{E}^{2/3}\sim 1$
in terms of the two smallest Lagrangian and Eulerian time scales, to
contrast with Ruelle's $\gamma \tau_{\eta}\sim 1$. In
  terms of the Taylor length's time scale $\tau_{\lambda}\equiv
  \lambda/(\varepsilon \lambda)^{1/3}$, it can be written as
  $\tau_{\gamma} \tau_{\lambda} \sim \tau_{\eta}^{2}$.

\begin{figure}
	\centering 
	\includegraphics[width=0.5\textwidth]{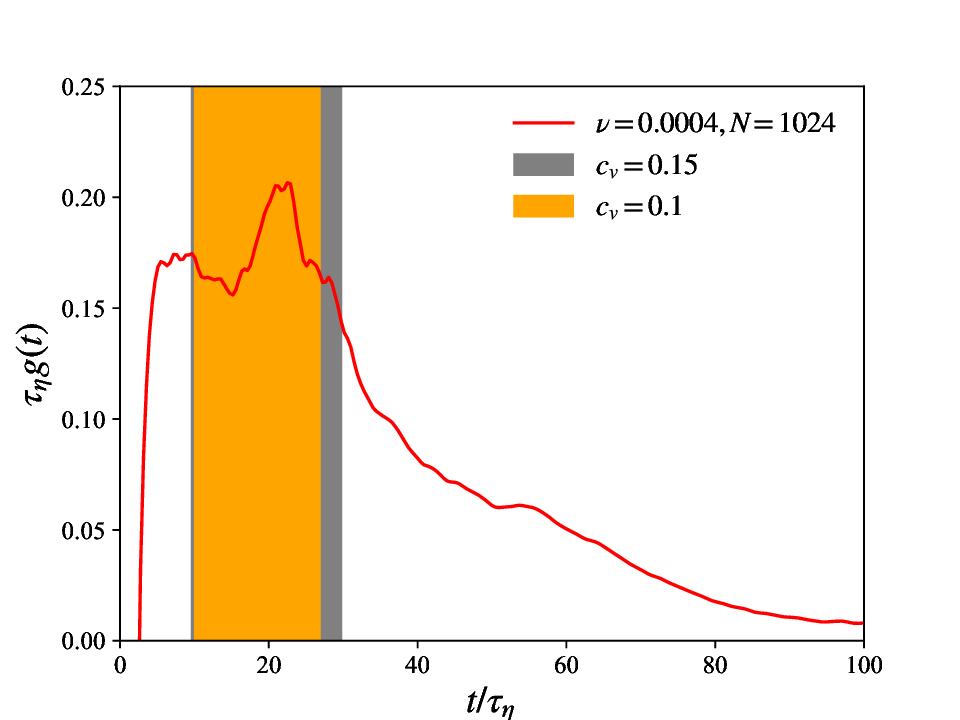}
	\caption{Example of $g(t)\equiv \frac{\left\langle
			E_{\Delta}(t+\Delta t)\right\rangle-\left\langle
			E_{\Delta}(t-\Delta t)\right\rangle}{4\Delta t\left\langle
			E_{\Delta}(t)\right\rangle}$ versus time $t$ ($\Delta t$
		is the sampling time interval). Both axes are
		non-dimensionalised by the reference field's Kolmogorov time
		scale. Two $c_v$ thresholds, 0.1 and 0.15, are applied to
		detect the plateau region of $g(t)$. Orange ($c_{v}=0.1$)
		and grey ($c_{v}=0.15$) colours are used to represent the
		plateau regions. For these two thresholds, the resulting
		time-average values of $g(t)$ over the detected plateaus
		(after continuous increase and before continuous decrease)
		are $5.87$ ($c_{v}=0.1$) and $5.81$ ($c_{v}=0.15$).}
	\label{fig:time evolution of the Lyapunov exponents} 
\end{figure}

It is important to appreciate that our approach does not only yield
the scaling (\ref{eq:Lyapunov exponent scaling}) of $\gamma$, it also
yields two entirely new and deeper relations which underpin this
scaling, the sweeping relation (\ref{eq:assumption advect 2 product
	2}) and the scaling (\ref{eq:LDelta}) of the uncertainty field's
integral length scale $L_{\Delta}$. It is therefore possible to
computationally test our approach by much more than just checking the
$Re_{\lambda}$ scaling of $\gamma$, i.e. by also checking the validity
of its two underpinning relations (\ref{eq:assumption advect 2 product
	2}) and (\ref{eq:LDelta}). In fact, recent direct numerical
simulations (DNS)
\cite{ge2023production,berera2018chaotic,boffetta2017chaos,mohan2017scaling},
have already shown that $\gamma\tau_{\eta}$ is not constant but
increases with Reynolds number. However, relations such as
(\ref{eq:assumption advect 2 product 2}) and (\ref{eq:LDelta}) have
never been proposed nor tested against data before.

\section{\label{sec:Simulations}Simulations}
 We validate our scaling laws (\ref{eq:assumption
	advect 2 product 2}), (\ref{eq:LDelta}) and (\ref{eq:Lyapunov
	exponent scaling}) with DNS of statistically stationary periodic
isotropic turbulence such as in
\cite{ge2023production,berera2018chaotic,boffetta2017chaos,mohan2017scaling}. We
integrate the incompressible Navier-Stokes equations
\begin{eqnarray}
	\label{eq:NS equation}
	\partial_{t}\bm{u}^{(i)}+\left(\bm{u}^{(i)}\cdot\bm{\nabla}\right)\bm{u}^{(i)}&=&-\bm{\nabla}p^{(i)}+\nu\bm{\nabla}^{2}\bm{u}^{(i)}+\bm{f}^{(i)},\nonumber \\ 
	\bm{\nabla}\cdot\bm{u}^{(i)}&=&0,
\end{eqnarray}
where $\bm{u}^{(1)} \equiv \bm{u}$ for $i=1$ and $\bm{u}^{(2)}=
\bm{u}+ \Delta\bm{u}$ for $i=2$. The governing equations are solved
numerically using a fully de-aliased pseudo-spectral code in a
periodic cube of length $2\pi$ with $N^{3}$ collocation points. We
simulate flows with a series of Reynolds numbers at two resolutions:
$k_{\max}\eta\approx1.7$ and $k_{\max}\eta\approx3.4$, where
$k_{\max}=N/3$ is the maximum resolvable wavenumber. Starting from a
statistically stationary velocity field $\bm{u}^{(1)}$, we generate
the perturbed velocity field $\bm{u}^{(2)}$ by randomly redistributing
the Fourier velocity modes at high wavenumbers $k \eta>1.5$ following
the method in \cite{ge2023production}. The parameters of our
simulations are reported in Table \ref{tab:main parameters}. The
present simulations correspond to case F2 in \cite{ge2023production}
where $\bm{f}^{(1)}= \bm{f}^{(2)}$ is a negative damping forcing
appplied to the largest length scales of the turbulence. The exact
form of the large-scale forcing has no impact on the evolution of
uncertainty during its chaotic growth because uncertainty is limited
to the smallest length scales of the turbulence during chaotic growth
(see \cite{ge2023production}).

We use Eq. (\ref{eq:exponential growth of uncertainty energy})
(effectively, the generalised Lyapunov exponent of
  order 2, see \cite{Cencinietal2010}) to determine $\gamma$. In
practice, we calculate the central difference $g(t)\equiv
\frac{\left\langle E_{\Delta}(t+\Delta t)\right\rangle-\left\langle
  E_{\Delta}(t-\Delta t)\right\rangle}{4\Delta t\left\langle
  E_{\Delta}(t)\right\rangle}$ (where $\Delta t$ is the sampling time
interval) at every time step and identify the time range where it is
oscillating around a constant value in time (see Fig. \ref{fig:time
  evolution of the Lyapunov exponents}). Times before this chaotic
time range are dominated by negative Lyapunov exponents and
dissipation of the uncertainty field \cite{ge2023production}
(continuous growth region in Fig. \ref{fig:time evolution of the
  Lyapunov exponents}); times after this chaotic time range
(continuous decrease region in Fig. \ref{fig:time evolution of the
  Lyapunov exponents}) are times when the uncertainty begins to
contaminate inertial range scales and its growth morphs from
exponential to power law (see
\cite{ge2023production,boffetta2017chaos,leith1971atmospheric}).

The chaotic time regime corresponding to the plateau region of $g(t)$
is detected by using the coefficient of variation $c_v$ of $g(t)$
($c_v$ is the ratio of the standard deviation to the time-average
value of $g$). We calculate $c_v$ for a sliding window over the entire
dataset of $g(t)$ with window size set to approximately $1/3$ of a
first visual estimate of the total plateau region
in plots such as Fig. \ref{fig:time evolution of the
    Lyapunov exponents} obtained from our various DNS runs. Regions
with $c_v$ below a threshold are identified as potential parts of the
overall plateau. Here this threshold is set to $0.1$. The mean value
of $g$ over each window is recorded as the window is sliding, the very
first window starting from $t=0$. Consecutive increases of this mean
value are not detected because our window is chosen to be large
enough. However, consecutive decreases of this mean value can be
detected and if they occur over a time range equal to twice the
sliding window size, then we are not in the plateau range we are
looking for. Alternatively, when the sliding mean of $g(t)$ does not
continously decrease over such a time range, then we take this time
range to be the chaotic regime's chaotic plateau.  It is possible that
excessive fluctuation magnitudes in some cases may result in multiple
disjoint plateau regions being identified. In such cases, the chaotic
time range of uncertainty is defined as the time from the start time
of the first plateau to the end time of the last plateau. We also
tried a $c_v$ threshold equal to 0.15, as shown in Fig. \ref{fig:time
  evolution of the Lyapunov exponents}, and the relative difference
between the mean values of $g(t)$
obtained with these two different thresholds is less than 0.05, which
is small compared to the values of $\gamma$ reported in Table
\ref{tab:main parameters}.

\section{\label{sec:Simulation results}Simulation results}
For all the Navier-Stokes turbulence cases
(see Table \ref{tab:main parameters}) where $k_{\max}\eta\approx1.7$
we have $\Delta t\lesssim0.1\gamma^{-1}$. For double-resolved cases
where $k_{\max}\eta\approx3.4$, the sampling time interval $\Delta t$
is half that of their corresponding $k_{\max}\eta\approx1.7$ case. As
can be seen in Table \ref{tab:main parameters}, this has no
significant effect on $\gamma$.
All data presented in Table \ref{tab:main parameters} and the
subsequent figures (notably $\gamma$, $L_{\Delta}$ and $Re_{\lambda}$)
are time-average values calculated by averaging over the time-interval
where $g(t)$
fluctuates in time around a constant. For the validation of
(\ref{eq:Lyapunov exponent scaling}) we use data from the present
letter's DNS but also data obtained by
\cite{berera2018chaotic,boffetta2017chaos,mohan2017scaling}
from their DNS of statistically stationary periodic isotropic
turbulence. However, the validation of the new scaling laws
(\ref{eq:assumption advect 2 product 2}) and (\ref{eq:LDelta}) relies
exclusively on the present letter's DNS data.

\begin{table*}
	\caption{\label{tab:main parameters}Parameters of the sixteen
          pairs (with/without uncertainty) of DNS of statistically
          stationary periodic Navier-Stokes turbulence run for this
          letter. With the exception of $N^3$, $\nu$ and $\epsilon_0$,
          all quantities in the table are time-averaged over the
          duration of the chaotic average exponential growth of
          uncertainty.  $\varepsilon_{0}$ is the turbulence
          dissipation rate preset by the negative damping forcing
          $\bm{f}^{(1)}= \bm{f}^{(2)}$ (see equation (3.7) and
          surrounding text in \cite{ge2023production} for full details
          on this forcing), $\varepsilon$ is the space-time-average
          turbulence dissipation rate,
          $U\equiv\left(\frac{2}{3}\left\langle\left|\bm{u}\right|^{2}/2\right\rangle\right)^{1/2}$,
          $L=(\pi/2U^{2})\int k^{-1}\hat{E}(k)\rm{d}$$k$ is the
          integral length scale of the reference field (in terms of
          its 3D energy spectrum $\hat{E}(k)$), $\gamma$ is obtained
          by averaging $0.5\left\langle
          E_{\Delta}\right\rangle^{-1}\mathrm{d}\left\langle
          E_{\Delta}\right\rangle/\mathrm{d}t$ over time during the
          exponential growth of uncertainty energy $\left\langle
          E_{\Delta}\right\rangle$, $Re\equiv UL/\nu$, and $k_{\max}$
          is the maximum resolvable wavenumber.}
	\begin{ruledtabular}
		\begin{tabular}{cccccccccccc}
				$N^{3}$&$\nu$&$\varepsilon_{0}$&$ \varepsilon$&$ U$&$L$&$ L_{\Delta}$&$\tau_{\eta}$&$\gamma$&$Re$&$Re_{\lambda}$&$ k_{\max}\eta$\\ \hline
		$2048^{3}$&$0.000157$&$0.10$&$0.087$&$0.625$&$1.149$&$0.016$&$0.042$&$5.852$&$4571$&$408.26$&$1.76$\\
		$1024^{3}$&$0.000400$&$0.10$&$0.110$&$0.605$&$0.938$&$0.038$&$0.060$&$2.937$&$1418$&$214.04$&$1.68$	\\
		$1024^{3}$&$0.000453$&$0.15$&$0.151$&$0.702$&$1.036$&$0.037$&$0.055$&$3.336$&$1605$&$230.93$&$1.70$	\\
		$1024^{3}$&$0.000498$&$0.20$&$0.218$&$0.772$&$0.986$&$0.037$&$0.048$&$4.080$&$1527$&$221.18$&$1.66$	\\	
		$1024^{3}$&$0.001000$&$0.10$&$0.099$&$0.638$&$1.176$&$0.094$&$0.101$&$1.287$&$751$&$158.64$&$3.42$	\\	
		$512^{3}$&$0.001000$&$0.10$&$0.091$&$0.614$&$1.130$&$0.094$&$0.105$&$1.192$&$693$&$153.49$&$1.74$	\\
		$512^{3}$&$0.001140$&$0.15$&$0.153$&$0.707$&$0.989$&$0.091$&$0.086$&$1.496$&$614$&$146.22$&$1.68$	\\
		$512^{3}$&$0.001260$&$0.20$&$0.162$&$0.770$&$1.196$&$0.098$&$0.088$&$1.436$&$731$&$161.17$&$1.79$	\\
		$512^{3}$&$0.002520$&$0.10$&$0.097$&$0.614$&$1.122$&$0.202$&$0.161$&$0.761$&$273$&$93.13$&$3.42$	\\
		$256^{3}$&$0.002520$&$0.10$&$0.102$&$0.611$&$1.094$&$0.201$&$0.157$&$0.762$&$265$&$90.17$&$1.69$	\\
		$256^{3}$&$0.002880$&$0.15$&$0.142$&$0.693$&$1.102$&$0.206$&$0.142$&$0.819$&$265$&$92.04$&$1.72$	\\
		$256^{3}$&$0.003170$&$0.20$&$0.201$&$0.789$&$1.170$&$0.197$&$0.126$&$0.930$&$291$&$95.51$&$1.70$	\\
		$256^{3}$&$0.006450$&$0.10$&$0.102$&$0.582$&$1.134$&$0.395$&$0.252$&$0.412$&$102$&$51.15$&$3.42$	\\
		$128^{3}$&$0.006450$&$0.10$&$0.102$&$0.578$&$1.133$&$0.415$&$0.252$&$0.435$&$101$&$50.49$&$1.69$	\\
		$128^{3}$&$0.007380$&$0.15$&$0.148$&$0.651$&$1.129$&$0.427$&$0.224$&$0.463$&$100$&$49.79$&$1.71$	\\
		$128^{3}$&$0.008130$&$0.20$&$0.199$&$0.728$&$1.160$&$0.413$&$0.202$&$0.546$&$104$&$51.12$&$1.70$	\\
		\end{tabular}
	\end{ruledtabular}
\end{table*}
\begin{figure}
	\centering 
	\includegraphics[width=0.5\textwidth]{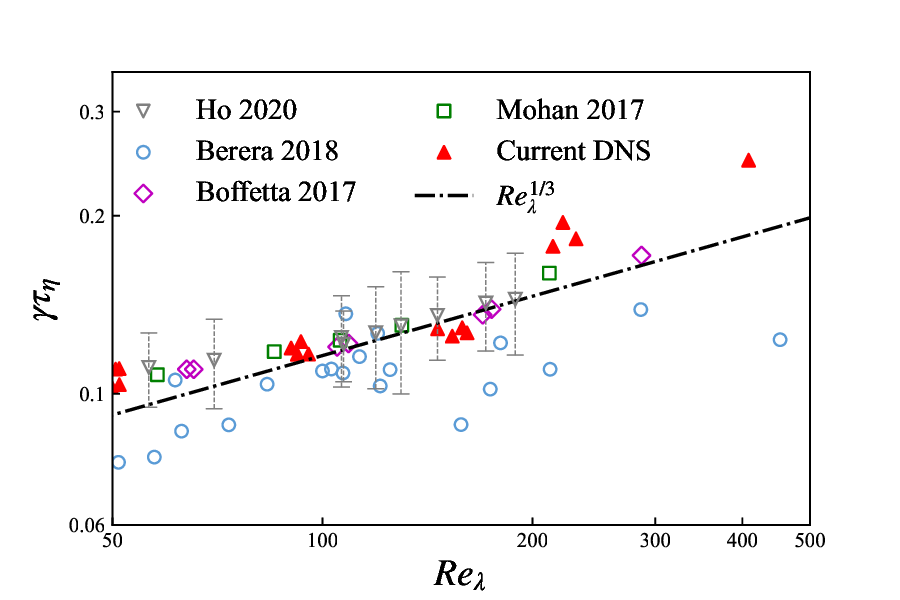}
	\caption{$Re_{\lambda}$ dependence of
          $\gamma\tau_{\eta}$. Results from present DNS (filled red
          triangles). Results from other DNS (hollow markers): blue
          circles \cite{berera2018chaotic}, magenta diamonds
          \cite{boffetta2017chaos}, green squares
          \cite{mohan2017scaling} and upside down triangles with error
          bars \cite{ho2020fluctuations}. The black dash-dotted line
          shows $Re_{\lambda}^{1/3}$ as per equation (\ref{eq:Lyapunov
            exponent scaling}).}
	\label{fig:GammaTaueta} 
\end{figure}

\begin{figure}
	\centering 
	\includegraphics[width=0.5\textwidth]{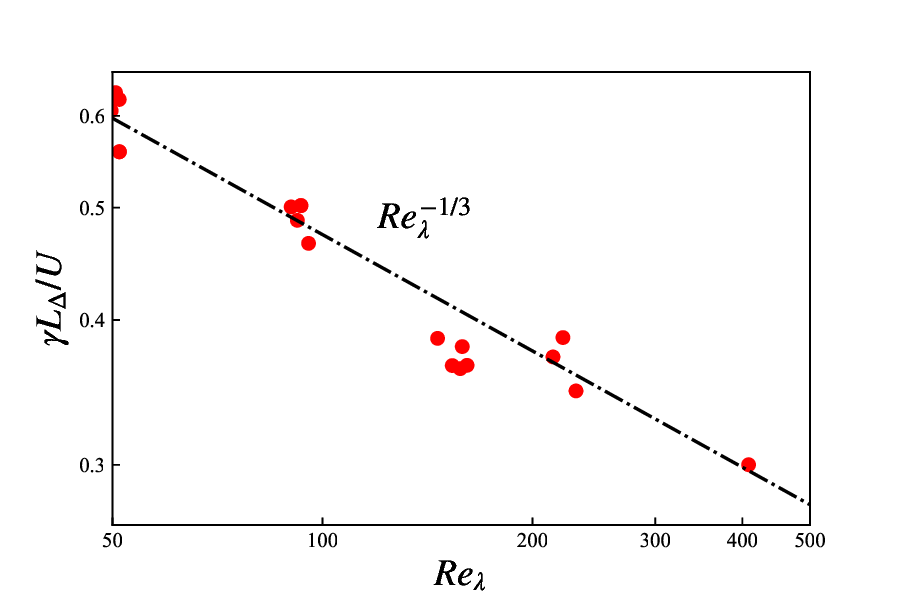}
	\caption{$Re_{\lambda}$ dependence of $\gamma
          L_{\Delta}/U$. The black dash-dotted line shows
          $Re_{\lambda}^{-1/3}$ as per sweeping equation
          (\ref{eq:assumption advect 2 product
            2}). The symbols are from the DNS run for
            this letter.}
	\label{fig:Gamma} 
\end{figure}

Fig. \ref{fig:GammaTaueta} shows a plot of $\gamma \tau_{\eta}$ versus
$Re_{\lambda}$. Our DNS data and those of previous DNS confirm that
the chaotic nature of uncertainty evolution is not characterised by
$\gamma\tau_{\eta}\sim1$ \cite{ruelle1979microscopic}. Instead,
Fig. \ref{fig:GammaTaueta} shows that $\gamma\tau_{\eta}\sim
Re_{\lambda}^{1/3}$ (Eq. (\ref{eq:Lyapunov exponent scaling})) is a
good fit of all the data (in particular ours and those of
\cite{boffetta2017chaos,mohan2017scaling,ho2020fluctuations}) even
though different methods for estimating $\gamma$ were used by
different authors. Note in particular the error bars
  in the data from \cite{ho2020fluctuations} who have shown that the
  relative standard deviation, $\sigma_{\gamma}/\gamma$, is
  approximately constant at $\sigma_{\gamma}/\gamma=0.2$ suggesting
  that the higher Reynolds number values of $\gamma$ in
  Fig. \ref{fig:GammaTaueta} are within reasonable range.

\begin{figure}
	\centering 
	\includegraphics[width=0.5\textwidth]{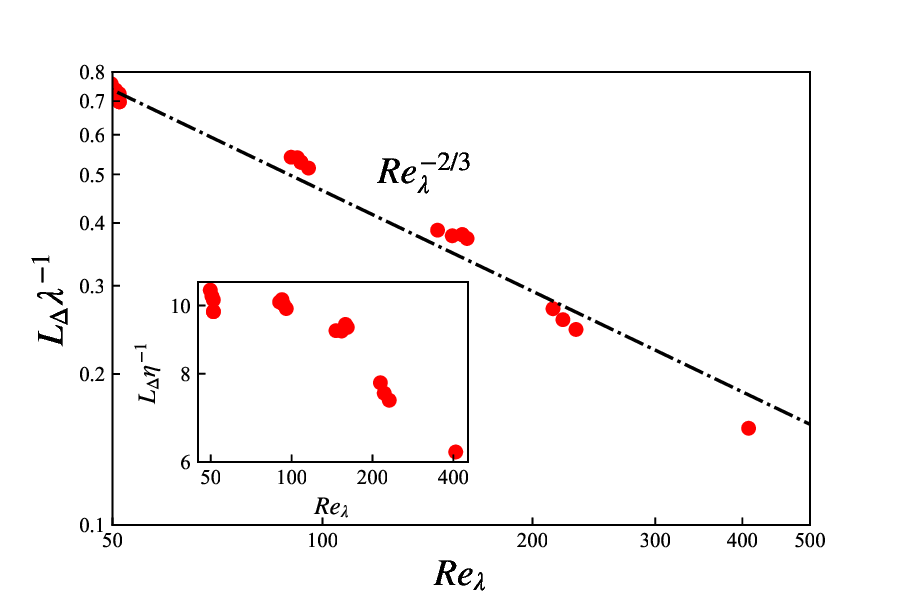}
	\caption{$Re_{\lambda}$ dependence of
		$L_{\Delta}\lambda^{-1}$. The black dash-dotted line shows
		$Re_{\lambda}^{-2/3}$ as per Eq. (\ref{eq:LDelta}). The
		inset shows $L_{\Delta}\eta^{-1}$ versus $Re_{\lambda}$ for
		the same data.}
	\label{fig:LDelta2Tay} 
\end{figure}

Figs. \ref{fig:Gamma} and \ref{fig:LDelta2Tay} present the
$Re_{\lambda}$ dependencies of $\gamma L_{\Delta}/U$ and
$L_{\Delta}\lambda^{-1}$ respectively. It is observed that both
$\gamma L_{\Delta}/U$ and $L_{\Delta}\lambda^{-1}$ depend on
$Re_{\lambda}$ as described by our scaling equations
(\ref{eq:assumption advect 2 product 2}) and (\ref{eq:LDelta}),
thereby supporting our theoretical approach. The scaling relation
(\ref{eq:LDelta}) implies $L_{\Delta}\eta^{-1}\sim
Re_{\lambda}^{-1/6}$, suggesting that uncertainty-containing scales
drop below the Kolmogorov length as $Re_{\lambda}$ increases to very
high values. The Reynolds numbers of our simulations are not high
enough to evidence values of $L_{\Delta}$ smaller than $\eta$, but our
DNS do show that $L_{\Delta}\eta^{-1}$ decreases with increasing
$Re_{\lambda}$ (inset of Fig. \ref{fig:LDelta2Tay}). Given that two
different spatial resolutions are applied to our DNS (see Table
\ref{tab:main parameters}) with almost identical results, spatial
discretization is an unlikely origin of the observed Reynolds number
scalings. The influence of time discretization has been excluded in
previous work \cite{mohan2017scaling}.

The Reynolds number scalings supported by the
  aforementioned three figures are validated over a range of Reynolds
  numbers from $Re_{\lambda}\approx 50$ to about $400$. It is worth
  noting that the power-law distribution $P(l)= Al^{-q}$ which
  underpins these scalings is supported (with $q=2/3$, even though the
  actual value of $q$ does not matter here) by experimental and
  numerical turbulence data for $Re_{\lambda}$ as low as $40$ (see
  \cite{ mazellier2008turbulence,ferran2023characterising} and note
  that the well-defined power law observed in these references is for
  a quantity that is mathematically equivalent to $P(l)= Al^{-q}$ by
  equation (2.4) in \cite{V91}).

\section{\label{sec:Conclusion}Conclusion}
We derived the $Re_{\lambda}$ scaling (\ref{eq:Lyapunov exponent
  scaling}) of the maximal Lyapunov exponent characterising the
chaotic exponential growth of small-scale (dissipation-range)
uncertainty in statistically stationary homogeneous isotropic
turbulence on the basis of (i) sweeping of uncertainty fluctuations by
Taylor length size turbulent eddies which leads to $\gamma
L_{\Delta}\sim URe_{\lambda}^{-1/3}$ and (ii) scalings of statistics
of consecutive zero-crossing distances which lead to
$L_{\Delta}\sim\lambda Re_{\lambda}^{-2/3}$. The new relations $\gamma
L_{\Delta}\sim URe_{\lambda}^{-1/3}$ and $L_{\Delta}\sim\lambda
Re_{\lambda}^{-2/3}$ are well-supported by our DNS data and imply
$\gamma\tau_{\eta}\sim Re_{\lambda}^{1/3}$ which aligns much better
with our and previous DNS results compared to $\gamma\tau_{\eta}\sim
1$ \cite{ruelle1979microscopic}. The scaling $L_{\Delta}/\lambda \sim
Re_{\lambda}^{-2/3}$ implies that the chaotic nature of very high
$Re_{\lambda}$ turbulence occurs at sub-Kolmogorov scales. It is known
that $\eta$ is an average inner length scale
(i.e. length scale where viscous and inertial forces
  balance) and that different inner length scales exist at different
locations in the turbulence, some of which are much smaller than the
average represented by $\eta$ (see
\cite{schumacher2007sub,frisch1995turbulence}).  Similarly, the
scaling $\gamma\tau_{\eta}\sim Re_{\lambda}^{1/3}$ implies that,
in the limit of growing Reynolds number, chaotic
growth of uncertainty happens with time scales smaller than
$\tau_{\eta}$ (the smallest Lagrangian time scale of the
turbulence). The sub-Kolmogorov nature of chaoticity implied by our
model for very high $Re_{\lambda}$ turbulence suggests that turbulence
is more unpredictable than previously thought.

\begin{acknowledgments}
Jin Ge acknowledges financial support from the China
Scholarship Council. We are grateful for access to Zeus supercomputers
(Mésocentre de Calcul Scientifique Intensif de l'Université de Lille).
\end{acknowledgments}


%
\end{document}